# An air-cooled Litz wire coil for measuring the high frequency hysteresis loops of magnetic samples – a useful setup for magnetic hyperthermia applications

V. Connord, B. Mehdaoui, R.P. Tan, J. Carrey* and M. Respaud

Laboratoire de Physique et Chimie des Nano-objets (LPCNO) ; Université de Toulouse; INSA; UPS; CNRS (UMR 5215) ; 135 avenue de Rangueil, F-31077 Toulouse, France

**Abstract :**

A low-cost and simple setup for measuring the high-frequency hysteresis loops of magnetic samples is described. An AMF in the range 6-100 kHz with amplitude up to 80 mT is produced by a Litz wire coil. The latter is air-cooled using a forced-air approach so no water flow is required to run the setup. High-frequency hysteresis loops are measured using a system of pick-up coils and numerical integration of signals. Reproducible measurements are obtained in the frequency range of 6-56 kHz. Measurement examples on ferrite cylinders and on iron oxide nanoparticle ferrofluids are shown. Comparison with other measurement methods of the hysteresis loop area (complex susceptibility, quasi-static hysteresis loops and calorific measurements) is provided and shows the coherency of the results obtained with this setup. This setup is well adapted to the magnetic characterization of colloidal solutions of MNPs for magnetic hyperthermia applications.

# Main Text:

I. Introduction

Magnetic hyperthermia has been the subject of an intense research activity in the past decade [1]. This experimental cancer therapy uses the heat generated by magnetic nanoparticles (MNPs) put in an alternating magnetic field (AMF) of typical frequency in the range 50-300 kHz. Experimentally, the heating power value is most of the time determined by measuring the temperature rise of a colloidal solution placed in AMF [2]. Since the heat generated by the MNPs during one cycle equals the area of their hysteresis loop [1], an alternative method consists in measuring directly the hysteresis loop. However, since the hysteresis loop shape depends tremendously on the frequency of the AMF, quasi-static measurements performed in a standard magnetometer are not satisfying; measurements should be done at a frequency similar to the one used in magnetic hyperthermia. Measuring the hysteresis loop instead of performing temperature measurements presents two major advantages: i) the complete hysteresis loop shape contains much more information than its simple area : information on saturation magnetization, magnetic interactions, aggregation of MNPs, MNP anisotropy can for instance be deduced from the hysteresis loop shape. We and other groups have already shown the interest of this method to get an insight into the physics of magnetic hyperthermia [3, 4, 5, 6]. ii) It is much faster than temperature measurements. A typical temperature measurement takes in itself around one minute, but the stabilisation of the temperature is much longer, so the typical delay between two measurements is ten minutes. A high-frequency hysteresis loop measurement takes in itself a few micro-seconds, with no need of waiting between two measurements.

So far, only a few groups have reported on the development of setups permitting such measurements. Several groups have reported on the building of susceptometers, which provide real and imaginary components of the susceptibility at low AMF [7]. A hysteresis loop tracer working at a moderate frequency of 2 kHz was reported in Ref. [8]. Bekovic *et al.* have built a setup with an objective and an approach similar to ours [6, 9]. However, because of the technology used for the production of the AMF, the maximum AMF amplitude was only 19 mT and the coil had to be water cooled. Finally, a setup with similar functionalities as the one which will be presented here has been built by Garaio *et al.* [10]. Unfortunately, very few technical and building details are provided in this reference, preventing anyone to envisage building a similar setup.

In this article, we present a setup permitting to measure the hysteresis loops of colloidal solutions of MNPs in a frequency range 6-56 kHz and up to 80 mT. Being based on the use of Litz wires, this setup presents the main advantage that a low power is necessary to produce the AMF so it can be air-cooled. Measurement results on typical samples are provided.

II. Description of the setup

1. Electric circuit

This setup is based on a resonant circuit similar to the one described in Ref. [2]. The principle is to produce an alternating current at a chosen frequency through a function generator (MTX 3240, Metrix) coupled to a voltage amplifier (HSA 4052, NP Corporation), which can deliver a current of ±2.8 A and voltage of ±150V at a maximum frequency of 500 kHz. The electrical circuit is shown in Fig. 1(a). Since the amplifier maximum current is limited to 2.8 A, a home-made transformer is used to increase the current amplitude. It is composed of 4 commercial I-shaped Ni-Zn ferrites assembled in square (Epsos, material N27). The transformer has 23 turns of Litz wire (240×0.05mm, Connect systemes) at the primary and 2.5 turns at the secondary, which increases the output current amplitude by a factor 9. To bring the transformer to resonance, a home-made high-voltage ceramic disks adjustable capacitance $C_1$ is introduced in the primary loop; its building has been described in Ref. [2]. A second capacitor $C_2$ is placed into the secondary circuit to bring it to resonance. $C_2$ is composed of several high-voltage ceramic capacitors in series (Vishay, 100 nF, 2500 V). The current in the setup is measured with an AC current probe (3274 clamp probe, Hioki). Finally, a home-made coil which is electrically equivalent to a LC parallel circuit is placed into the secondary circuit to produce the AMF.

2. Production of magnetic field by the main coil

The main magnetic coil is composed of Litz wires (480x0.071mm, Pack Feindrähte). Using Litz wire is essential to avoid the skin effect due to the high frequency current, and to keep the impedance of the coil as low as possible. 120 turns of wires are rolled up around a PVC structure. Each layer of wire is separated from the others by a 0.5 mm thick fiberglass mesh to prevent electrical breakdown and sparking between the layers. Thanks to the low impedance of the coil, the heat generated inside it is moderate and can be extracted by a forced air approach. For that purpose, the coil former is pierced of several rectangular holes allowing the air to go trough [see Fig. 1(b)]. The coil is then put on a hollow holder and connected to a simple vacuum cleaner (Ultra Active Green, Electrolux) so the air is forced from outside the coil to the inside [see Figs 2(a) and 2(b)]. This cheap system keeps the coil temperature stable even if a high amplitude AMF is applied during hours. AMF amplitude was calibrated using a pick up coil inserted into the main coil. The amplitude of the AMF is calculated using :

$$\mu_0 H_{max} = \frac{\varepsilon}{nS_{coil}\pi f}, \qquad (1)$$

where $n$ and $S_{coil}$ are the number of turns and surface of the pick-up coil, $f$ the frequency of the AMF and $\varepsilon$ the voltage appearing at the coil terminal. Fig. 3(a) shows the AMF maximum amplitude as a function of the coil current at a fixed frequency of 54 kHz. AMF amplitude is linear with the applied current, with a factor of 2.0±0.1 mT/A. Fig 3(b) shows the evolution of the AMF amplitude as a function of the position inside the coil. As expected, the field amplitude decreases on the sides of the coil. There is a 4 cm depth plateau where AMF does not vary by more than 4%. This 4 cm height zone is used as a working area to put the sample and the measurement coils (see below).

Table 1 displays the evolution of the maximum AMF amplitude and current which can be generated as a function of the working frequency. The corresponding coil impedance $Z(\Omega)$, as well as the $C_1$ and $C_2$ values leading to the circuit resonance are also shown. Coil impedance increases with the applied frequency inducing a decrease of the maximum AMF. Simultaneously, this also increases the heat generated inside the coil, which is compensated by adjusting the vacuum cleaner power.

3. Hysteresis loop measurements

Now the home-made system permitting to measure the high-frequency hysteresis loop of magnetic samples is described. The detection system is schematized in Fig. 4(a). It consists of two identical contrariwise-wounded pick-up coils (7 turns of a copper wire, 0.7 mm diameter) connected in series. Let us call coil 2 the pick-up coil wounded around the sample and coil 1 the other one [see Fig. 1(c)]. These pick-up coils are wounded around a PVC holder which maintains the sample and the pick-up coils at a constant height inside the main coil [see Fig. 1(c)]. A tapping and a nut at the top of the sample holder permit to adjust precisely its height. Two signals are required to measure the high-frequency hysteresis loops: let $e_1(t)$ be the voltage at the terminals of pick-up coil 1 and $e_2(t)$ the one at the terminals of the two coils in series [see Fig. 4(a)]. These high-frequency signals are measured by an oscilloscope (TDS 2022B, Tektronix) connected by USB to a computer and then transmitted to the latter.

The protocol to measure the hysteresis loop of the sample is the following. First, the height of the empty sample holder is adjusted roughly to get a maximized signal in coil 1. Then a fine adjustment of the height is done by minimizing $e_2(t)$ signal. Each hysteresis cycle measurement then requires three steps:

- Measurement of $e_1(t)$ and $e_2(t)$ for a blank sample, which is the same vessel as the true sample filled with the same quantity of solvent but without any magnetic material.
- Measurement of the true sample
- Then, the signal $e_2(t)$ from the blank sample is substracted from the signal $e_2(t)$ obtained from the true sample.

A typical signal obtained from coil 1 is shown in Fig. 4(b). Signals $e_2(t)$ obtained from a magnetic sample and from the blank sample are shown in Fig. 4(c). To obtain the magnetic field and magnetization values, $e_1(t)$ and $e_2(t)$ are integrated numerically using :

$$\mu_0 H(t) = \frac{\int e_1(t)dt}{nS_{coil}} \quad (2)$$

$$\sigma(t) = \frac{\int e_2(t)dt}{\mu_0 n\rho S_{sample}\Phi} \quad (3)$$

$$M(t) = \frac{\int e_2(t)dt}{\mu_0 n S_{sample} \Phi} \qquad (4)$$

$\sigma$ is the magnetization per unit mass of magnetic material, $M$ its magnetization per unit volume, $S_{sample}$ the surface of a section of the magnetic sample, $\Phi$ the volume concentration of the sample and $\rho$ the density of the magnetic material. The numerical integration is performed on data coming from a single period of the AMF; the average value of the signal on this period is substracted from the signal before integration. The hysteresis loop is obtained by plotting $\sigma$ or $M$ as a function of $\mu_0 H$, as shown in Fig. 4(d).

### III. Measurement examples.

#### 1. Ferrite cylinders

To validate our setup, we have performed measurements on commercial ferrite cylinders. In Fig. 5, measurements of ferrites cylinders (Ferroxcube ROD8/2563S3, 8 mm diameter) at fixed frequency as a function of AMF amplitude are shown. Raw signals in coils 1 and 2 are shown in Figs. 5(a) and 5(b) for a 2.5 cm long ferrite cylinder. Corresponding hysteresis loops obtained after numerical calculation are shown in Fig. 5(c). These ferrite cylinders display a negligible coercive field, which explains why the various hysteresis loops all collapse on a single reversible curve. The ferrite saturation is clearly observed. On this experiment, the hysteresis loop could not be measured at larger field because $e_2(t)$ became too large to be measured by the oscilloscope. The approximate saturation value ($\approx 0.25$ T) matches the one expected for this ferrite (0.32 T from constructor). A low AMF, the cycles display a linear part, the slope of which corresponds to the external magnetic susceptibility $\chi_{ext}$, which in this case equals 10.5.

External susceptibility $\chi_{ext}$ is directly linked to the demagnetizing factor of the cylinder, itself being related to the cylinder length. To check the influence this parameter, we have measured the response of ferrite cylinder of different lengths [see Fig. 5(d)]. As expected, $\chi_{ext}$ strongly diminishes when shorter cylinders are measured. Three different theoretical values of $\chi_{ext}$ have been calculated. $\chi_m$ and $\chi_f$ are derived from the magnetometric and fluxmetric demagnetizing factors extracted from the tables of Ref. [11]. $\chi_{Boz}$ is another calculation derived from fluxmetric demagnetizing factors by R. M. Bozorth [12]. Comparison between these theoretical values and the experimental are shown in Table 2. The agreement between the experimental value and the theoretical one is acceptable for short ferrites, since there is a factor of 2 between both. The discrepancy increases significantly for longer samples, probably because long samples have a significant part out of the working area and are thus submitted in average to a lower AMF. However, both the reversible hysteresis loops and the saturation value expected for these ferrite cylinders are first signs of our setup validity.

2. Measurements on an iron oxide nanoparticle ferrofluid.

a) Hysteresis loop measurements.

The samples studied in this work were colloids of magnetite/maghemite prepared by a modified version of the well known co-precipitation method originally due to Kalafallah and Reimers [13]. The samples were prepared by precipitation of the oxyhydroxides from molar solutions of ferrous ($Fe^{2+}$) and ferric ($Fe^{3+}$) salts. The precipitation was undertaken using ammonia. Following the initial precipitation gentle warming was used to convert the oxyhydroxides nominally to magnetite ($Fe_3O_4$) but due to the alkalinity of the solution partial oxidation to maghemite ($Fe_2O_3$) occurred. Subsequently using closely controlled conditions of temperature and pH a controlled growth process (CGP) was used to produce a system with a narrow particle size distribution. The particle size distribution was measured using a JEOL 2011 TEM with a resolution of about 0.3 nm. Particle sizes were measured using a Zeiss particle size analyser which is essentially a light box such that the diameter of individual particles is obtained by an equivalent circle method. To ensure good statistics over 500 particles were measured and as expected a good fit to a lognormal distribution function was found. Figs. 6(a) and 6(b) show a transmission electron micrograph of the particles and the subsequent size distribution. The particles were dispersed in water using DMSA at a concentration of 5 mg of Fe per ml of water. As can be seen from the TEM image the particles were relatively well dispersed with little aggregation. The colloid was dialysed to remove remaining traces of the initial salt solutions.

To measure the sample, we put 0.5 mL of colloidal solution inside a 5 mm in diameter vessel. The measurement process here is exactly the same as for the ferrite samples except that, at the end of the measurement, Equ. (3) is used instead of Equ.(4) so the magnetization per mass of MNPs $\sigma$ is obtained. In Fig. 7 hysteresis loops measurements as a function of the AMF amplitude and for four frequencies in the range of our setup (19, 32, 56 and 92 kHz) are shown. For each frequency, all the cycles are plotted on the same graph. Measurements at the three lower frequencies show a very high coherency, all the curves being interlocked one inside the others. At the largest frequency (92 kHz), the observed lack of coherency reflects a lack of reproducibility on the measurements. This is due to the fact that, at large frequencies, the signal is more sensitive to even slight modification of the position of the sample inside the sample holder. As a consequence, when this setup was used for measuring various nanoparticle systems, we have always restricted our frequency to a maximum value of 56 kHz [3, 4, 5].

b) Comparison between our setup and other measurement methods.

To check the validity and coherency of the results obtained using our setup, we have compared it to three other measurement approaches. First, at low AMF, any magnetic system responds linearly with the applied magnetic field so its magnetic response is completely characterized by its complex susceptibility $\tilde{\chi}$. In this regime, the hysteresis loop is an ellipse, the area of which can be calculated using [1]:

$$A = \pi H_{max}^2 \chi'', \qquad (5)$$

where $\chi''$ is the imaginary component of $\tilde{\chi}$. In Fig. 8(a), we show that the hysteresis loops measured between 2 and 30 mT at 56 kHz are all ellipses with the same shape once normalized. The data are perfectly fitted with an ellipse using Equs. (30) and (31) of Ref. 1. From this fit, $|\tilde{\chi}|$, $\varphi$ (the phase delay between the AMF and the magnetization) and so $\chi'' = |\tilde{\chi}| \sin \varphi$ were determined. The hysteresis area calculated using Equ.(5) and $\chi''$ determined this way are shown in Fig. 8(b) along with the area resulting from the integration of the individual hysteresis loops; both are logically in good agreement. In this regime where the linear response theory is valid, standard ac susceptibility measurements are in principle sufficient to determine the hysteresis area. To check it, we have connected the output of the pick-up coils to a lock-in amplifier (SR830 DSP, Stanford Research Design). As expected, phase values obtained from the lock-in amplifier are independent of the AMF amplitude in the range 2-23 mT, leading to a phase value of $\varphi = 8.7 \pm 2.6°$. This means that at low magnetic, the results given by our setup matches what would be obtained with a standard susceptometer.

Second, we have compared the hysteresis loop obtained on our setup with the one obtained using a vibrating sample magnetometer (VSM) measuring the static hysteresis loop. For that purpose, a dried powder issued from the ferrofluid was measured. In Fig. 8(c), both measurements are compared. The static hysteresis loop measured at the VSM has a negligible coercive field whereas the high-frequency one displays an opened hysteresis loop. In spite of this non-surprising difference due to the frequency dependence of the coercive field, we notice that the amplitude of the magnetization and the curvature of the hysteresis loop are very similar in both setups, which is another sign of the validity of our setup.

Finally, we compare the obtained hysteresis loop with a calorific method. Indeed, the hysteresis area is related to the specific absorption rate (SAR) of MNPs by equation [1]:

$$SAR = Af. \qquad (6)$$

We have thus performed SAR measurements using the protocol and analysis method described in Ref. [2]. Briefly, it consists in measuring the temperature rise of the colloidal solution when the AMF is put on. We have performed these measurements i) directly inside the present setup using the sample holder as a calorimeter, and ii) inside the electromagnet described in Ref. [2]. The hysteresis area deduced from these temperature measurements are plotted in Fig. 8(d) along with the values obtained after integrating the hysteresis loops. Hysteresis area deduced from temperature measurements on the electromagnet are in very good agreement with the one obtained from integration, which is a last confirmation of the validity of our setup. However, it is obvious from data shown in Fig. 8(d) that the present setup is not adapted to perform temperature measurements; it is very likely that, due to the presence of a strong air flow inside the setup to cool down the coil, the calorimeter losses are very large and prevent to perform any correct calorimetric measurements inside the setup. This point could be improved by inserting an adiabatic chamber inside the sample holder to insulate thermally the sample from the remaining of the setup.

IV. Conclusion

We designed a Litz wire coil able to generate an AMF in the range 6-100 kHz with amplitude up to 80 mT. This coil is air-cooled so no water flow is required to run the setup. Magnetic hysteresis loops are obtained using contrary-wide wounded pick-up coils inserted in a coil space where the AMF is homogeneous. Pick-up coils signals are acquired by an oscilloscope and then numerically integrated. Reproducible and stable hysteresis loops are obtained up to frequency of 56 kHz. At low AMF, when the system responds linearly to the AMF, connecting the coils to a lock-in simply transforms the setup into a standard susceptometer. In these conditions, hysteresis area analysis and complex susceptibility measurements give identical results with respect to the magnetic response of the system. Comparison with VSM and temperature measurements indicates the coherency of the our measurement results. The present setup permits an insight on the physics of magnetic hyperthermia and has proven its utility in previously published articles [3, 4, 5].

**Acknowledgements :**

This research was partly funded by the European Community's Seventh Framework Programm under grant agreement no. 262943 "MULTIFUN". We acknowledge Liquids Research for supplying the magnetic nanoparticles through the MULTIFUN Project. We thank A. Khalfaoui and B. Simonigh for machining the setup.

**References :**

* Correspondence should be addressed to J.C. (julian.carrey@insa-toulouse.fr)

**Tables:**

| $f$ (kHz) | $C_1$ (nF) | $C_2$ (nF) | $I$ (A) | $Z$ (Ω) | $\mu_0 H_{max}$ (mT) |
|---|---|---|---|---|---|
| 6.8 | 2200 | 1100 | 40 | 0.25 | 80 |
| 13.5 | 657 | 560 | 40 | 0.2 | 75 |
| 22.5 | 200 | 200 | 40 | 0.2 | 75 |
| 31.9 | 111 | 100 | 40 | 0.28 | 75 |
| 45.0 | 53.1 | 50 | 40 | 0.33 | 75 |
| 55.2 | 34.1 | 33 | 40 | 0.4 | 75 |
| 63.7 | 24 | 25 | 35 | 0.45 | 65 |
| 71.1 | 20 | 20 | 33 | 0.5 | 61 |
| 84.1 | 7.8 | 14.3 | 28 | 0.6 | 52 |
| 95.4 | 4 | 11.1 | 25 | 0.63 | 46 |

Table 1: Electrical and magnetic properties of the main coil as a function of the working frequency $f$. Values of capacitor in the primary and secondary circuit ($C_1$ and $C_2$), of the maximum current $I$, of the impedance $Z$ and of the magnetic field $\mu_0 H_{max}$ are provided.

| Length (mm) | $\chi_{ext}$ | $\chi_f$ | $\chi_m$ | $\chi_{Boz}$ |
|---|---|---|---|---|
| 6.25 | 1.9 | 4.329 | 4 | 4.35 |
| 12.5 | 3.9 | 12.48 | 7.87 | 9.09 |
| 25 | 8.1 | 31.25 | 17.24 | 25 |
| 50 | 19.2 | 104.17 | 35.71 | 66.7 |
| 75 | 22.1 | 177.3 | 71.43 | 833 |

Table 2: Comparison between the external value of the susceptibility $\chi_{ext}$ and different theoretical values. $\chi_{Boz}$ is extracted from Ref. [12]. $\chi_f$ and $\chi_m$ are extracted from Ref. [11]. The constructor value for the ferrite magnetic permeability $\mu = 350$ was used.

**Figures :**

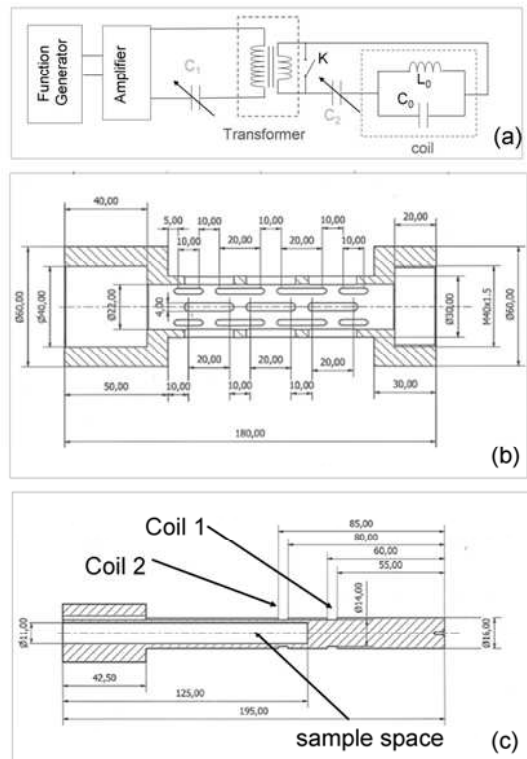

Figure 1 (color online) : (a) Electric schematic of the setup. $C_1$ ($C_2$) is a variable capacitor permitting the resonance of the primary (secondary) circuit. The main coil is schematized by a parallel $L_0C_0$ circuit. (b) Technical drawing of the main coil holder. (c) Technical drawing of the sample holder.

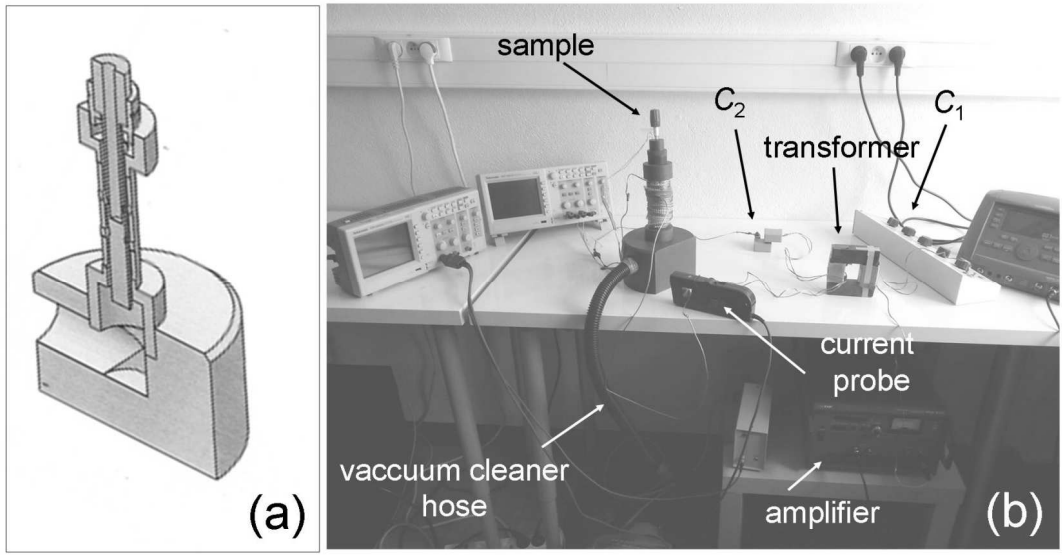

Figure 2 (color online): (a) Technical drawing of the complete setup (b) Picture of the complete setup.

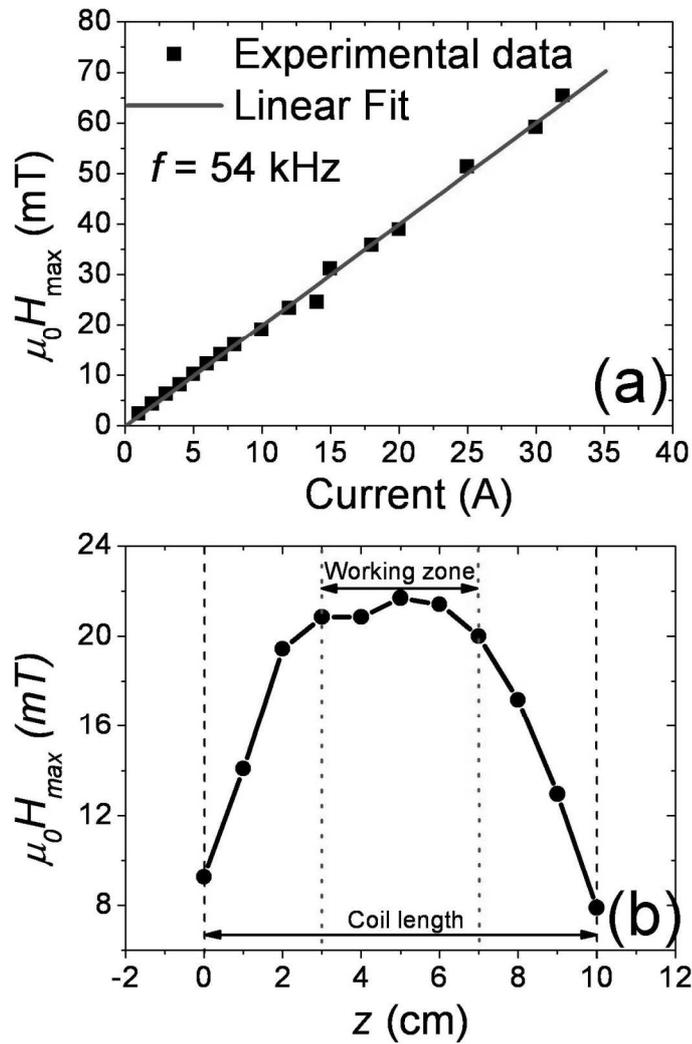

Figure 3 (color online): (a) AMF measured as a function of the ac current sent through the main coil. (b) AMF amplitude as a function of the position inside the coil. Vertical dashed lines show the limits of the coil. Dotted lines delimitate the working zone, where the pick-up coils and the sample are placed.

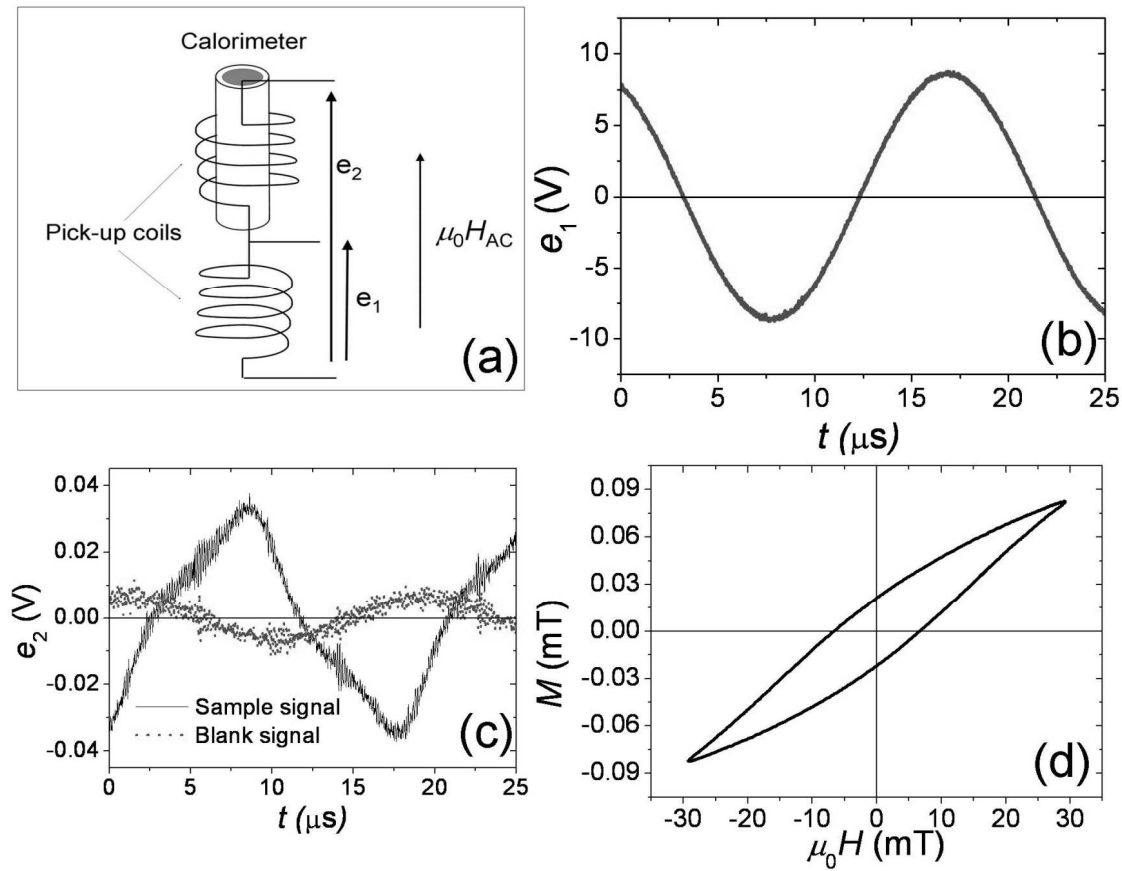

Figure 4 (color online): (a) Schematic illustrating the detection principle using pick-up coils. (b) Typical measurement of voltage signal $e_1(t)$ measured on coil 1. (c) Typical measurement of voltage signal $e_2(t)$ measured on coil 1 and coil 2 in series. Dots correspond to the signal obtained when a blank sample is inserted. Plain line corresponds to the signal obtained when a typical magnetic sample is put inside the setup. (d) Hysteresis loop obtained after subtracting the blank sample signal from the sample one, and subsequent numerical integration.

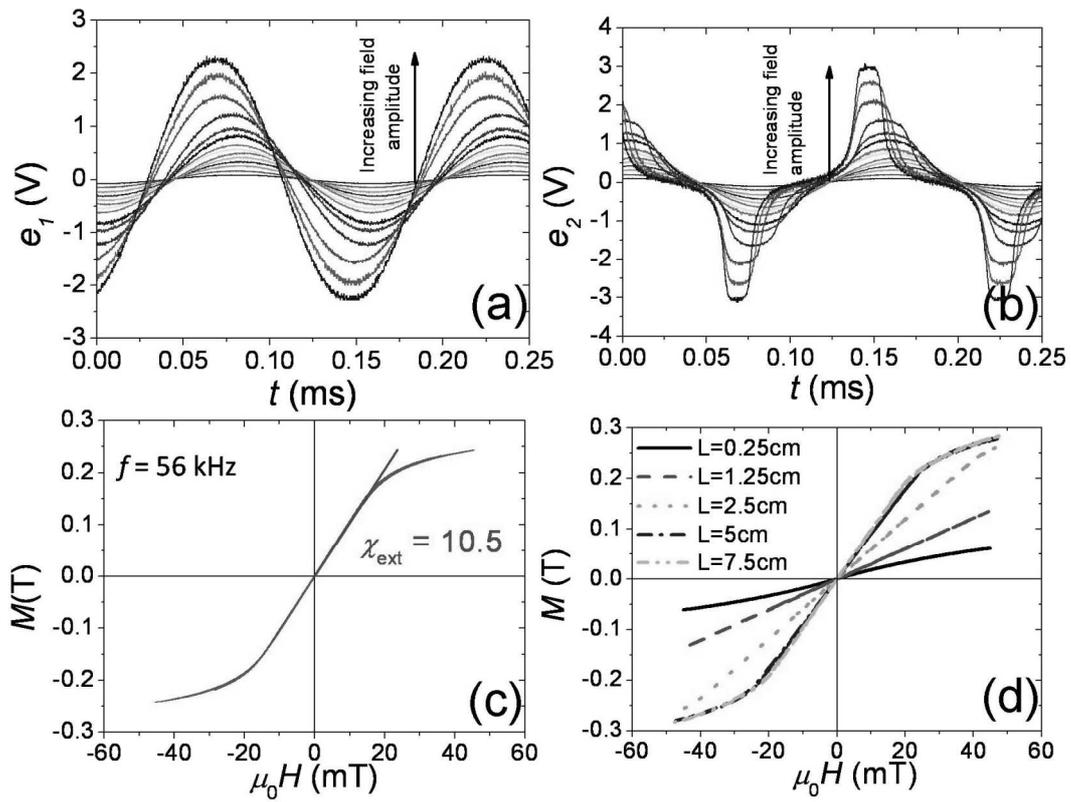

Figure 5 (color online): Measurements of ferrite cylinder samples. (a)(b) Signals $e_1(t)$ and $e_2(t)$ when the AMF is increase from 0 to 50 mT at a frequency of 56 kHz. A 2.5 cm long ferrite sample is measured (c) Hysteresis loops obtained from the previous signals. All curves all superimposed on a single one. (d) Hysteresis loops obtained for ferrite samples of varying length $L$.

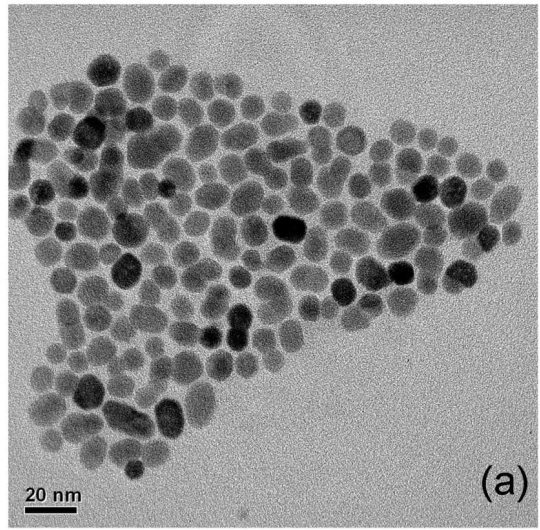

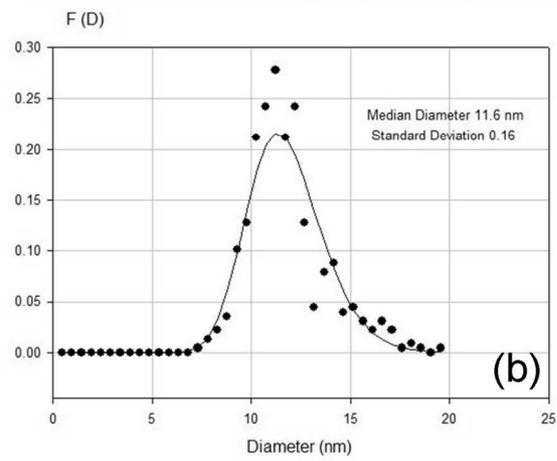

Figure 6 : (a) Transmission electron microscopy of the synthesized MNPs. (b) Size distribution of the MNPs fitted by a log-normal distribution.

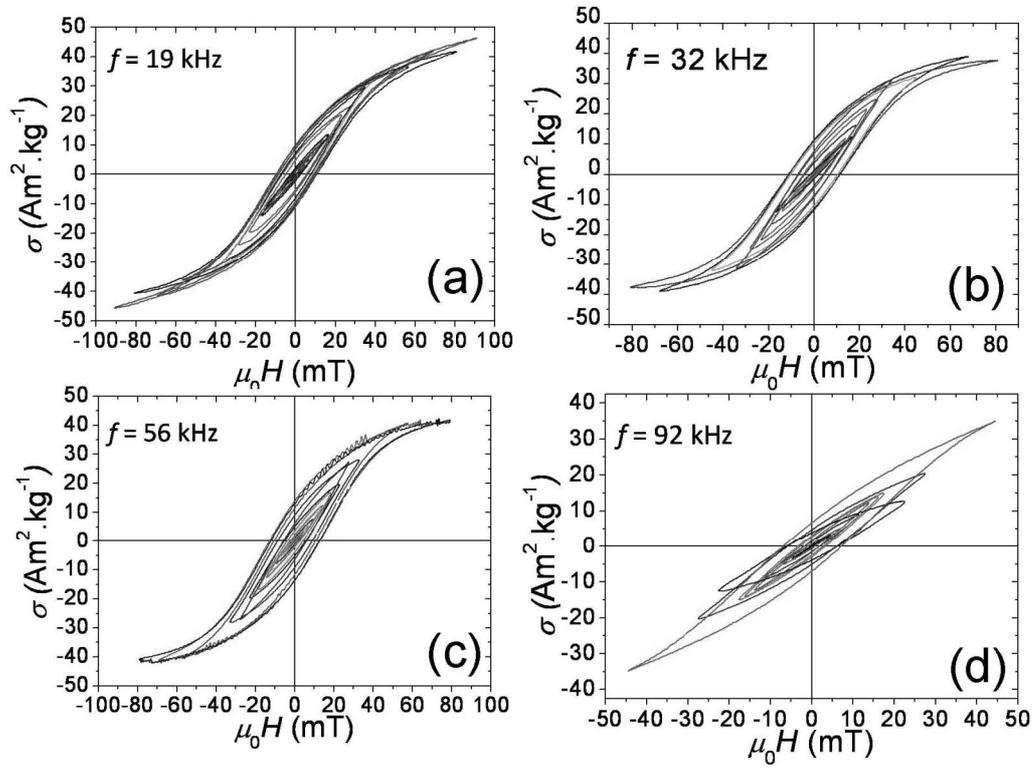

Figure 7 (color online): Measurements on an iron oxide nanoparticle ferrofluid. The figures show the hysteresis measurements as a function of the AMF at frequencies of a) 19 kHz, b) 32 kHz, c) 56 kHz, and d) 92 kHz.

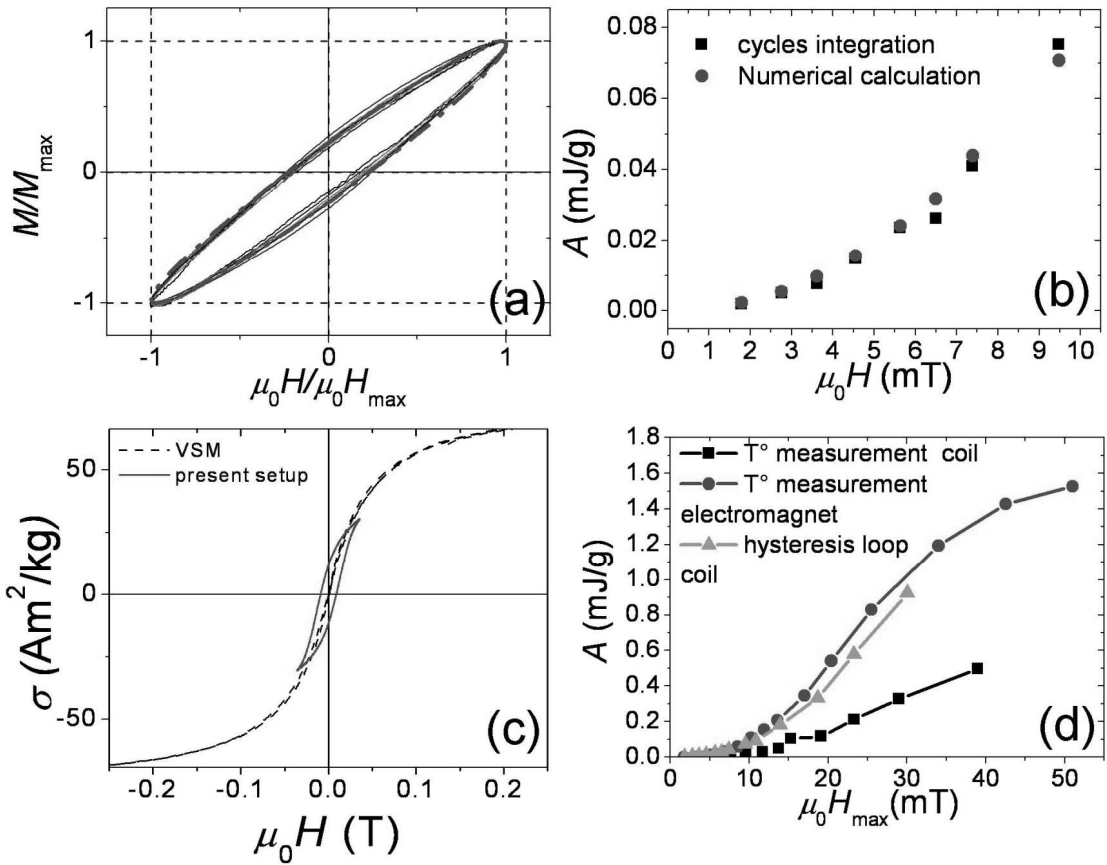

Figure 8 (color online): Comparison between various measurement methods. (a) Normalized hysteresis loops measured at 56 kHz for $\mu_0H_{max}$ ranging from 2 to 18 mT. (b) Area calculated at low AMF by two methods: (■) by integration of the hysteresis loop. (●) by using Equation (5). (c) Comparison between the hysteresis loop measured using our setup (plain line) and VSM (dashed line). (d) The hysteresis area is calculated using (■) temperature measurements inside our setup, (●) temperature measurement inside an electromagnet, (▲) integration of hysteresis loops.